# Second Harmonic Generation of MoSi$_2$N$_4$ Layer

Lei Kang* and Zheshuai Lin†

*Technical Institute of Physics and Chemistry, Chinese Academy of Sciences, Beijing 100190, China*
E-mails: kanglei@mail.ipc.ac.cn (L.K.); zslin@mail.ipc.ac.cn (Z.L.)

**The recently discovered two-dimensional (2D) layered semiconductor MoSi$_2$N$_4$ has aroused great interest due to its unique 2D material characteristics. In this Letter, we found that differences in the structural details for MoSi$_2$N$_4$ may lead to differences in the intensity of second harmonic generation (SHG) and its response to strain. Accordingly, SHG can be used as a simple technique to identify the structural details of this system. We further calculated the SHG effects of MoSi$_2$N$_4$ derivatives and investigated their strain-regulation mechanism, especially including the anomalous SHG responses under strain for MoSi$_2$P$_4$ and MoGe$_2$P$_4$, differing from other known 2D materials. The studies may have forward-looking significance for the research of nonlinear optics and optoelectronics in this novel 2D material system.**

Two-dimensional (2D) layered semiconductors are very important in the application of nanoscale optoelectronic devices [1-6]. $MoSi_2N_4$ is a newly discovered 2D material system with rich optoelectronic properties and has attracted great scientific attention since its discovery [7]. In previous studies, a little structural detail of the 2D $MoSi_2N_4$ layer is still unclear. Although density functional theory (DFT) calculations have locked in possible forms, more structural information and evidence are still needed. Fortunately, this transition metal nitride has a similar Mo-N structural symmetry ($D_{3h}$) as Mo-S in monolayer (ML) $MoS_2$, which is an important 2D material with a macroscopic second-order optical nonlinearity that can be detected by second harmonic generation (SHG) technique [8,9]. If the specific structural forms of $MoSi_2N_4$ are sensitive to SHG response, the actual atomic structure can be well identified according to the intensity or change of SHG. More importantly, SHG performance is very useful for optoelectronics (*e.g.*, valleytronics) in 2D materials [10-14]. The origin and regulation (*e.g.*, under strain) of SHG may be of interest in nonlinear optical (NLO) physics, which has received extensive attention in the study of material properties [15-20]. Therefore, it is urgent to study the SHG characteristics of $MoSi_2N_4$ system.

In this Letter, we studied the SHG effects of typical $MoSi_2N_4$ phases, analyzed the SHG difference in combination with SH polarization (*P*), and simulated the SHG response to strain from first principles. Theoretical results indicate that the two most typical phases may exhibit different SHG intensities due to the difference in *P* between the outer and inner layers of the sandwich structure. Moreover, in the analogous $MoSi_2P_4$ and $MoGe_2P_4$ systems, the SHG difference becomes more extraordinary, exhibiting an SHG effect comparable to ML $MoS_2$. Remarkably, due to the different coordination forms of the outer tetrahedral and inner octahedral layered structures, their SH polarization responses to strain are also different, resulting in different changes in the total SHG response to strain. In particular, the two phases of $MoSi_2P_4$ and $MoGe_2P_4$ show opposite SHG regulation under strain, originated from the anomalous effect of charge transfer in the outer Si/Ge-P layers, which are different from other known 2D NLO semiconductors.

The first-principles calculations are performed by the plane-wave pseudopotential method based on the DFT using CASTEP [21,22]. The norm-conserving pseudopotentials, energy cutoff of 770 eV and Monkhorst-Pack *k*-point meshes of 0.04 Å$^{-3}$ in the Brillouin zone are adopted, respectively [23]. The structures are fully relaxed using the BFGS scheme [24]. The electronic structures and optical properties are calculated by standard DFT method with the PBESOL functional [25,26], which is an efficient way to accurately calculate the SHG properties of NLO materials [27-29]. The static SHG coefficients $\chi^{\alpha\beta\gamma}$ are calculated using the following expression:

$$\chi^{\alpha\beta\gamma} = \chi^{\alpha\beta\gamma}(VE) + \chi^{\alpha\beta\gamma}(VH)$$

where $\chi^{\alpha\beta\gamma}(VE)$ and $\chi^{\alpha\beta\gamma}(VH)$ denote the contributions from virtual-electron (VE) processes and virtual-hole (VH) processes, respectively. The formulas can be written as follows [30]:

$$\chi^{\alpha\beta\gamma}(VE) = \frac{e^3}{2\hbar^2 m^3} \times \sum_{vcc'} \int \frac{d^3k}{4\pi^3} P(\alpha\beta\gamma) \, Im(p_{vc}^{\alpha} p_{cc'}^{\beta} p_{c'v}^{\gamma}) \left( \frac{1}{\omega_{cv}^3 \omega_{vc'}^2} + \frac{2}{\omega_{vc}^4 \omega_{c'v}} \right)$$

$$\chi^{\alpha\beta\gamma}(VH) = \frac{e^3}{2\hbar^2 m^3} \times \sum_{vv'c} \int \frac{d^3k}{4\pi^3} P(\alpha\beta\gamma) \, Im(p_{vv'}^{\alpha} p_{v'c}^{\beta} p_{cv}^{\gamma}) \left( \frac{1}{\omega_{cv}^3 \omega_{v'c}^2} + \frac{2}{\omega_{vc}^4 \omega_{cv'}} \right)$$

Here, $\alpha$, $\beta$ and $\gamma$ are Cartesian components; *v* and *v'* denote valence bands (VBs); *c* and *c'* denote conduction bands (CBs); $P(\alpha\beta\gamma)$ denotes full permutation and explicitly shows the Kleinman symmetry of the SHG coefficients. The band energy difference and momentum matrix elements are denoting as $\hbar\omega_{ij}$ and $p_{ij}^{\alpha}$, respectively.

Note that ML $MoS_2$ can be served as a benchmark for 2D NLO materials with similar $E_g$. According to the definition of 2D SHG coefficient $\chi(2D)$, it is equal to the corresponding 3D SHG coefficient multiplied by the structure thickness (unit is Å·pm/V) [31]. Using the $MoS_2$ benchmark, we can define the relative SHG coefficient $\chi_R(2D)$ to describe the 2D SHG capability as:

$$\chi_R(2D) = \chi(2D)/\chi(MoS_2)$$

For comparison, the predicted results of the lattice constants, energy bandgaps and SHG coefficients for $MoS_2$ are listed in **Table S1** of the Supporting Information, which is highly consistent with the available data [9], indicating the accuracy of our computational method.

Combining the experimental characterization with previous DFT studies [7], the two most possible structural forms of $MoSi_2N_4$ (namely α-phase and β-phase) are shown in **Fig. 1**. They both consist of an inner $MoS_2$-like $MoN_2$ layer and two outer buckled Si-N layers, which are stacked together by interlayer Si-N bonds to form a sandwich layered structures. To qualitatively compare the charge distribution and covalent strength, atomic and bond Mulliken population (MP) are calculated from first principles as listed in **Table S2** [32]. MP analysis shows that due to different hybrid forms and coordination environments, the inner N atoms ($N_i$, charge MP ≈ -0.9) and the outer N atoms ($N_o$, charge MP ≈ -1.3) show a large difference in charge, resulting in a significant bond difference between interlayer Si-$N_i$ (bond MP ≈ 0.5) and outer Si-$N_o$ (bond MP ≈ 2.0). The bond MP are mainly distributed on the inner Mo-$N_i$ bonds (bond MP ≈ 1.2) and the outer Si-$N_o$ bonds, indicating relatively strong covalency for the inner $N_i$-Mo-$N_i$ layer and the outer Si-$N_o$ layers. In addition, since the charge transfer between Mo and $N_i$ is smaller than that between Si and $N_o$, the orbital-coupling gap of Mo-$N_i$ should be smaller than that of Si-$N_o$, which would mainly determine the energy bandgap $E_g$ of $MoSi_2N_4$. In the above sense, α-$MoSi_2N_4$ and β-$MoSi_2N_4$ have similar chemical bond structures and total (formation) energies, which are difficult to be distinguished by the difference in lattice constant and energy bandgap.

**Figure 1.** Structures of α-$MoSi_2N_4$ (a) and β-$MoSi_2N_4$ (b), and the SH polarization (*P*) of layers.

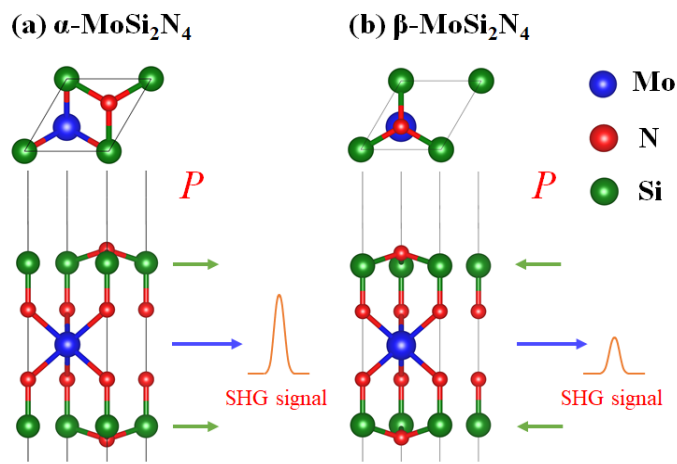

It should be emphasized that the distribution of positive and negative charges in the sandwich structure is different, which is induced by different sites of surface $N_o$ anions. For α-$MoSi_2N_4$

(β-MoSi$_2$N$_4$), the alignment of SH polarization (*P*) in the inner layer is same (opposite) to that of the outer layer (**Fig. 1**). As a result, for α-MoSi$_2$N$_4$ (β-MoSi$_2$N$_4$), the *P*-induced SHG effects for inner and outer layer are additive (cancelled), so the total SHG signals of the two phases are different. Moreover, the SH polarization mainly comes from the joint determination of local dipole moment and bond covalency [33], so the SHG effect of the inner Mo-N$_i$ layer is stronger than that of the outer Si-N$_o$ layer. Therefore, the total SHG signal of α-MoSi$_2$N$_4$ should be greater than that of β-MoSi$_2$N$_4$. This result is consistent with our initial speculation that the SHG can identify subtle structural differences if the SHG responses are different. **Table S1** lists the calculated SHG coefficients $\chi_{111}=\chi_{122}$ for α-MoSi$_2$N$_4$ and β-MoSi$_2$N$_4$, with 0.58 and 0.49 times of ML MoS$_2$, respectively, which is basically consistent with our analysis on SH polarization in **Fig.1**.

**Fig. 2** further plots the partial density of states (PDOS) and band-resolved SHG density of α-MoSi$_2$N$_4$ and β-MoSi$_2$N$_4$ [34]. The orbitals of inner Mo-N$_i$, outer Si-N$_o$ and interlayer Si-N$_i$ exhibit basically consistent PDOS distribution, which means strong hybridization among them. The outer Si-N$_o$ orbitals show a wider energy gap than the inner Mo-N$_i$ orbitals, which is consistent with the MP analysis of charge transfer. The electronic states around $E_g$, especially on the top of VB, are mainly derived from Mo-N$_i$ bonds and Mo$^{4+}$ nonbonding states, which are sensitive to external photoelectric field and play a dominate role in SH polarization. The simulated SHG-weighted density in **Fig. 2** indeed shows that the SHG response mainly comes from the inner Mo-N$_i$ orbitals. The band-resolved SHG analysis further proves that although the orbitals near band edge look similar [34], the corresponding SH polarizations are different, resulting in a negative impact on the SHG response of some orbitals in β-MoSi$_2$N$_4$ (**Fig. 2**), which makes the total SHG intensity of β-MoSi$_2$N$_4$ being smaller than that of α-MoSi$_2$N$_4$.

**Figure 2.** PDOS and band-resolved SHG analysis of α-MoSi$_2$N$_4$ (a) and β-MoSi$_2$N$_4$ (b). Inset is the SHG-weight density projected to each atom.

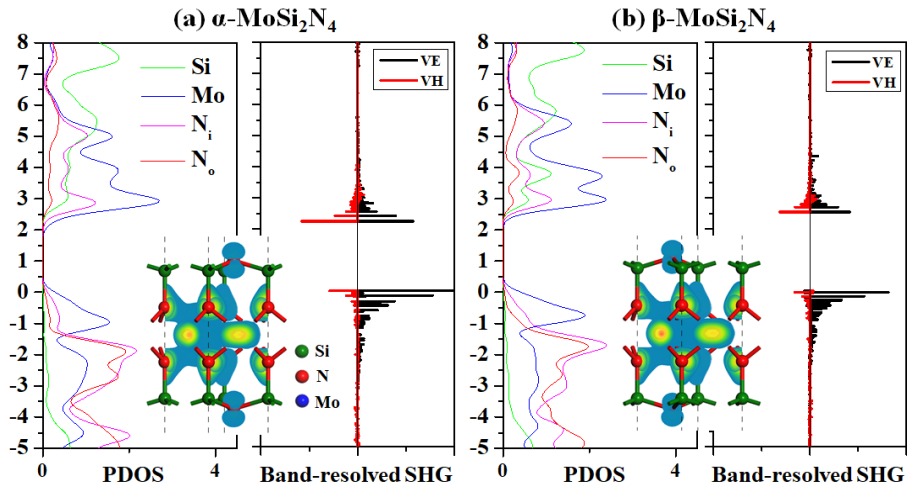

However, the SHG difference between α-MoSi$_2$N$_4$ ($\chi_{111} \approx 0.58\times$MoS$_2$) and β-MoSi$_2$N$_4$ ($\chi_{111} \approx 0.49\times$MoS$_2$) is too small to be distinguished easily in experiments, which is mainly because the outer Si-N$_o$ induced *P* is much smaller than the inner Mo-N$_i$ induced *P*. Theoretically, the magnitude of *P* mainly depends on the length of chemical bond and the distribution of anisotropic charges. From the MP analysis in **Table S2**, it can be found that inner Mo-N$_i$ layer with smaller $E_g$ has longer bond length than the outer Si-N$_o$ layer, resulting in stronger SH polarization based on the formula of SHG coefficient [30]. We also designed two hypothetical layered structures, i.e., MoN$_2$ and SiN by hydrogen passivated. The calculated results as listed in **Table S3** demonstrate that the H-passivated MoN$_2$ layer exhibits smaller $E_g$ (~ 2.2 eV) and much larger $\chi_{111}$ (~ 0.48×MoS$_2$) than those of H-passivated SiN layer ($E_g \approx$ 4.0 eV, $\chi_{111} \approx$ 0.04×MoS$_2$), in good agreement with the theoretical understanding although temporarily ignoring the interlayer Si-N$_i$ hopping.

To further distinguish the SHG difference between α-phase and β-phase, the SHG properties and MP results of typical MoSi$_2$N$_4$ derivatives, including WSi$_2$N$_4$, MoGe$_2$N$_4$, MoSi$_2$P$_4$ and MoGe$_2$P$_4$ with α-phase and β-phase, are studied as listed in **Table S1** and **Table S2**. Specifically, compared with α-MoSi$_2$N$_4$, since the charge MP (~ -0.94) of N$_i$ in Si-N$_i$ is larger and the W-N$_i$ bond length (~ 2.087 Å) is shorter, α-WSi$_2$N$_4$ has a larger $E_g$ (~ 2.3 eV) and smaller $\chi_{111}$ (~ 0.30×MoS$_2$) so the SHG difference changes small; Since the charge MP (~ -0.83) of N$_i$ in Ge-N$_i$ is smaller and the Mo-N$_i$ bond length (~ 2.117 Å) is close, the $E_g$ (~ 2.0 eV) of α-MoGe$_2$N$_4$ is smaller and $\chi_{111}$ (~ 0.60×MoS$_2$) is similar, so the SHG difference is still not large. As comparison, α-MoSi$_2$P$_4$ and α-MoGe$_2$P$_4$ exhibit much smaller $E_g$ (~ 0.86 and 0.82 eV) and much larger $\chi_{111}$ (~ 2.12 and 2.23×MoS$_2$) due to smaller charge MP (~ -0.12 and -0.09) of P$_i$ in Si/Ge-P$_i$ and longer Mo-P$_i$ bond length (~ 2.442 and 2.445 Å), so the SHG difference is sufficiently enlarged and can be distinguished easily. The calculated results of β-phases show similar trends.

**Figure 3.** Relative changes of the SHG coefficient $\chi_{111}$ with respect to strain ε from -3% to 3% for MoSi$_2$N$_4$ (a), MoSi$_2$P$_4$ (b) and MoGe$_2$P$_4$ (c), as compared with ML MoS$_2$

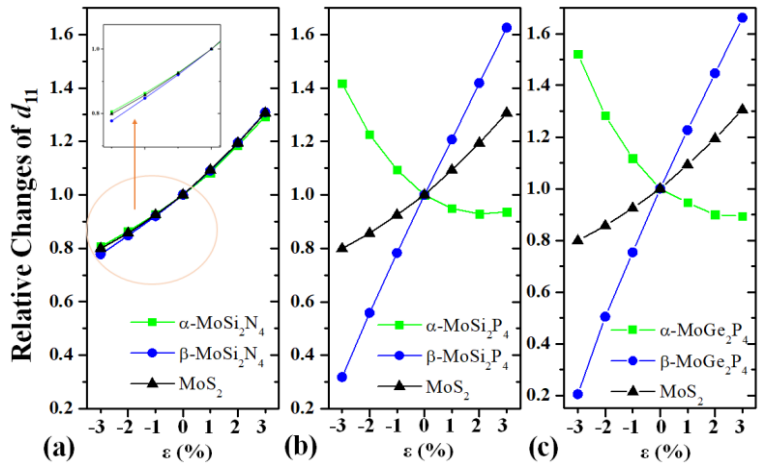

In fact, in addition to the SHG intensity, the change of SHG response to strain can be used to clearly identify the details of structural differences. As such, we further investigated the strain-dependent SHG effects in MoSi$_2$N$_4$ and MoSi$_2$P$_4$ as two representatives to compare with that of MoS$_2$. Note that different coordination polyhedral structures generally have different responses to strain. Accordingly, the inner Mo-N$_i$ octahedra may exhibit different SHG changes as compared with the outer Si-No tetrahedra. This can be seen from the strain-dependent SHG coefficients of hypothetical hydrogen-passivated MoN$_2$ and SiN layers (**Table S3**). Unfortunately, as shown in **Fig. 3a,** the SHG changes for α-MoSi$_2$N$_4$ and β-MoSi$_2$N$_4$ are very similar, slightly larger and smaller than that of MoS$_2$. As for α-MoSi$_2$P$_4$ and β-MoSi$_2$P$_4$, the SHG responses with respect to strain are completely different, as plotted in **Fig. 3b,** respectively exhibiting the trend of negative and positive correlation. This is an obvious difference and can be verified through experiments.

It is worth noting that in $MoSi_2P_4$, the reverse arrangement of SH polarization in the outer layer can cause the SHG effect to respond differently to strain. Unlike most 2D NLO materials, this is the abnormal SHG response to strain. The reason for this abnormal response is that, firstly, the outer Si-$P_o$ polarization contribute a lot to the SHG effect, so its change under strain would affect the total change of the SHG response. Secondly, the tetrahedral hybrid strength of Si-P is weaker than that of Si-N, so P atoms are less likely to be exposed on the surface since too many dangling bonds would result in more instability. As the strain varies from ε to -ε, the outer Si-$P_o$-Si angle becomes smaller and the dangling bonds on the P atoms would increase. As a result, the dangling electronic states on the outer $P_o$ atoms need to be transferred onto the adjacent Si atoms to maintain the stability of the sandwich structure. This is the opposite of the situation in $MoSi_2N_4$, as illustrated in **Fig. 4a** (for details see the MP analysis in **Table S4**). This anomalous charge transfer that occurs in the outer layered structure enlarges the covalency of the outer Si-$P_o$ bonds and enhances the SH polarization. This can also be seen from the SHG density distribution in **Fig. 4b**. Note that the SHG density is mainly distributed on the inner Mo-$P_i$ and outer Si-$P_o$ layer. When the strain changes from 3% to -3%, the SHG density on the outer Si-$P_o$ layer does not decrease but increases, indicating that the SHG effect increases as the strain decreases. This anomalous SHG increase with decreasing strain makes the total SHG effects for the α-phase and β-phase have a great contrast under the regulation of strain. A similar contrast also appears in the $MoGe_2P_4$ structure (**Fig. 3c**) because the charge transfer on the outer Ge-$P_o$ layer has a similar anomalous effect (**Table S4**).

**Figure 4.** Schematic diagram of charge transfer for α-$MoSi_2N_4$ and α-$MoSi_2P_4$ from strain ε to -ε (a), and SHG density of α-$MoSi_2P_4$ when strain ε is -3% and 3% (b).

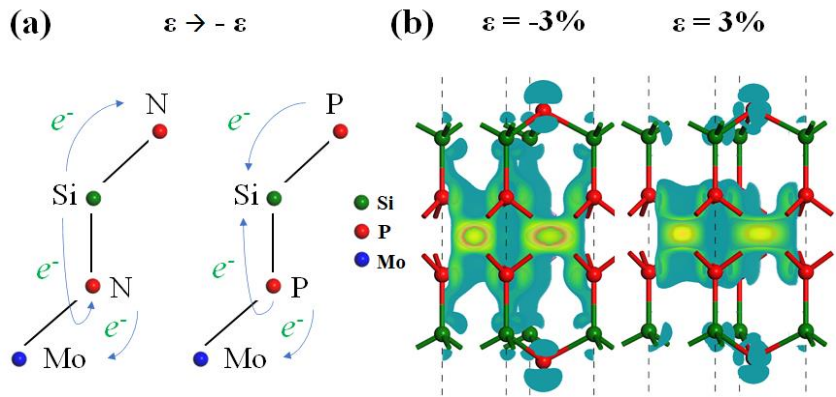

In summary, we have studied the SHG effects and strain-regulation mechanism of $MoSi_2N_4$ and its derivatives, which demonstrate that the SHG responses and changes with respect to strain can be used to identify detailed structural information of the $MoSi_2N_4$ system. Particularly, the anomalous SHG responses to strain were observed in $MoSi_2P_4$ and $MoGe_2P_4$, differing from other known 2D NLO materials, which are mainly derived from the anomalous effect of charge transfer on the Si/Ge-P layers. This study may arouse the interests in NLO applications and strain regulation of this novel 2D material system.

**Acknowledgements:** This work was supported by the National Natural Science Foundation of China (Grant No. 11704023).

**References:**

1. K. S. Novoselov, A. Mishchenko, A. Carvalho and A. H. Castro Neto, *Science*, 2016, **353**.
2. Q. Tang and Z. Zhou, *Prog. Mater. Sci.*, 2013, **58**, 1244-1315.
3. L. K. Li, Y. J. Yu, G. J. Ye, Q. Q. Ge, X. D. Ou, H. Wu, D. L. Feng, X. H. Chen and Y. B. Zhang, *Nature Nanotechnology*, 2014, **9**, 372-377.
4. B. Peng, P. K. Ang and K. P. Loh, *Nano Today*, 2015, **10**, 128-137.
5. A. Carvalho, M. Wang, X. Zhu, A. S. Rodin, H. B. Su and A. H. C. Neto, *Nat. Rev. Mater.*, 2016, **1**, 16.
6. X. Liu, Q. Guo and J. Qiu, *Adv. Mater.*, 2017, **29**.
7. Y. L. Hong, Z. B. Liu, L. Wang, T. Y. Zhou, W. Ma, C. Xu, S. Feng, L. Chen, M. L. Chen, D. M. Sun, X. Q. Chen, H. M. Cheng and W. C. Ren, *Science*, 2020, **369**, 670-+.
8. N. Kumar, S. Najmaei, Q. Cui, F. Ceballos, P. Ajayan, J. Lou and H. Zhao, *Phys. Rev. B*, 2013, **87**, 6.
9. Y. Li, Y. Rao, K. Mak, Y. You, S. Wang, C. R. Dean and T. F. Heinz, *Nano Letters*, 2013, **13**, 3329-3333.
10. M. Fiebig, V. V. Pavlov and R. V. Pisarev, *J. Opt. Soc. Am. B-Opt. Phys.*, 2005, **22**, 96-118.
11. M. Merano, *Opt. Lett.*, 2016, **41**, 187-190.
12. A. Autere, H. Jussila, Y. Dai, Y. Wang, H. Lipsanen and Z. Sun, *Adv. Mater.*, 2018, **30**.
13. Y. Wang, J. Xiao, S. Yang, Y. Wang and X. Zhang, *Opt. Mater. Express*, 2019, **9**, 1136-1149.
14. H. Chu, C. J. Roh, J. O. Island, C. Li, S. Lee, J. Chen, J.-G. Park, A. F. Young, J. S. Lee and D. Hsieh, *Phys. Rev. Lett.*, 2020, **124**.
15. X. Yin, Z. Ye, D. A. Chenet, Y. Ye, K. O'Brien, J. C. Hone and X. Zhang, *Science*, 2014, **344**, 488-490.
16. J. Lee, K. F. Mak and J. Shan, *Nature Nanotechnology*, 2016, **11**, 421-+.
17. L. Wu, S. Patankar, T. Morimoto, N. L. Nair, E. Thewalt, A. Little, J. G. Analytis, J. E. Moore and J. Orenstein, *Nature Physics*, 2017, **13**, 350-355.
18. Z. Y. Sun, Y. Yi, T. C. Song, G. Clark, B. Huang, Y. W. Shan, S. Wu, D. Huang, C. L. Gao, Z. H. Chen, M. McGuire, T. Cao, D. Xiao, W. T. Liu, W. Yao, X. D. Xu and S. W. Wu, *Nature*, 2019, **572**, 497-+.
19. J. Liang, J. Zhang, Z. Z. Li, H. Hong, J. H. Wang, Z. H. Zhang, X. Zhou, R. X. Qiao, J. Y. Xu, P. Gao, Z. R. Liu, Z. F. Liu, Z. P. Sun, S. Meng, K. H. Liu and D. P. Yu, *Nano Letters*, 2017, **17**, 7539-7543.
20. S. H. Rhim, Y. S. Kim and A. J. Freeman, *Applied Physics Letters*, 2015, **107**, 5.
21. M. D. Segall, P. J. D. Lindan, M. J. Probert, C. J. Pickard, P. J. Hasnip, S. J. Clark and M. C. Payne, *J. Phys.-Conden. Matter*, 2002, **14**, 2717-2744.
22. S. J. Clark, M. D. Segall, C. J. Pickard, P. J. Hasnip, M. J. Probert, K. Refson and M. C. Payne, *Zeitschrift Fur Kristallographie*, 2005, **220**, 567-570.
23. A. M. Rappe, K. M. Rabe, E. Kaxiras and J. D. Joannopoulos, *Physical Review B*, 1990, **41**, 1227-1230.
24. B. G. Pfrommer, M. Cote, S. G. Louie and M. L. Cohen, *J. Comput. Phys.*, 1997, **131**, 233-240.
25. J. P. Perdew, K. Burke and M. Ernzerhof, *Phys. Rev. Lett.*, 1996, **77**, 3865-3868.
26. J. P. Perdew, A. Ruzsinszky, G. I. Csonka, O. A. Vydrov, G. E. Scuseria, L. A. Constantin, X. L. Zhou and K. Burke, *Phys. Rev. Lett.*, 2008, **100**, 4.
27. J. L. P. Hughes, Y. Wang and J. E. Sipe, *Phys. Rev. B*, 1997, **55**, 13630-13640.
28. L. Kang, F. Liang, X. Jiang, Z. Lin and C. Chen, *Acc. Chem. Res.*, 2020, **53**, 209-217.
29. S. Zhang, L. Kang and Z. Lin, *Nanoscale*, 2020, **12**, 14895-14902.
30. J. Lin, M. H. Lee, Z. P. Liu, C. T. Chen and C. J. Pickard, *Phys. Rev. B*, 1999, **60**, 13380-13389.
31. H. Wang and X. F. Qian, *Nano Letters*, 2017, **17**, 5027-5034.
32. I. Mayer, *Int. J. Quantum Chem.*, 1984, **26**, 151-154.
33. B. F. Levine, *Phys. Rev. Lett.*, 1970, **25**, 440-&.
34. M. H. Lee, C. H. Yang and J. H. Jan, *Phys. Rev. B*, 2004, **70**, 11.